\documentclass[twocolumn,showpacs,superscriptaddress,prl,aps]{revtex4}
\usepackage{graphicx}
\usepackage{amsmath}
\usepackage{amssymb}
\usepackage{latexsym}
\usepackage{array}
\usepackage{amsfonts}
\usepackage{mathrsfs}
\usepackage{verbatim}

\begin{document}

\title{Quantum Learning Machine}

\author{Jeongho Bang}

\affiliation{Department of Physics, Hanyang University, Seoul 133-791, Korea}

\author{James Lim}

\affiliation{Department of Physics, Hanyang University, Seoul 133-791, Korea}

\author{M. S. Kim}

\affiliation{School of Mathematics and Physics, The Queen's University of
Belfast, BT7 1NN, United Kingdom}

\author{Jinhyoung Lee}

\affiliation{Department of Physics, Hanyang University, Seoul 133-791, Korea}

\received{\today}

\begin{abstract}
  We propose a novel notion of a quantum learning machine for
  automatically controlling quantum coherence and for developing quantum
  algorithms. A quantum learning machine can be trained to learn a
  certain task with no {\em a priori} knowledge on its algorithm. As an example, it is demonstrated that the quantum learning machine learns Deutsch's task
  and finds itself a quantum algorithm, that is different from but
  equivalent to the original one.
\end{abstract}

\maketitle

\newcommand{\bra}[1]{\left<#1\right|}
\newcommand{\ket}[1]{\left|#1\right>}
\newcommand{\abs}[1]{\left|#1\right|}
\newcommand{\expt}[1]{\left<#1\right>}
\newcommand{\braket}[2]{\left<{#1}|{#2}\right>}
\newcommand{\commt}[2]{\left[{#1},{#2}\right]}

{\em Quantum automatic control}.-
Quantum information science (QIS) aims to exploit quantum mechanics to improve the acquisition, transmission and processing of information.  This field has seen explosive growth in recent years, stimulated by the applications such as quantum cryptography and quantum communication which have the potential to surpass their classical counterparts.  In particular, quantum computation which was originally proposed by Feynman~\cite{feynman}, received its momentum after considerable speedup was found for some algorithms including the Deutsch-Josza~\cite{dj_a1,dj_a2}, Shor's factorization~\cite{a_qa2-Shor, a_qa4-SG}, Grover's database search~\cite{a_qa3-Grover}, and hidden subgroup problem algorithms.  One of the important challenges in quantum computation is to find algorithms which can fully explore quantum parallelism for speedup.

For the success of QIS, developing new quantum algorithms and enhancing the controllability of quantum coherence and the ability for quantum-state engineering~\cite{b_qinfo} are important agendas and deserve novel approaches. Here, we propose a new method of quantum control, where quantum-state engineering is ``automatically'' implemented by using a feed-back method to adjust unitary operations so to eventually bring about a target state intended. The feed-back system performs a single-shot quantum measurement on the output system and figures out if it is in the target state. If not, it modifies the control parameters of unitary operation. The feed-back system repeats over an ensemble of given quantum systems, one by one, until the adjusted unitary
operation outputs the target state.

In fact, various feed-back systems have been studied for classical and quantum automatic controls which include the quantum neural network~\cite{a_nn}, quantum-state estimation~\cite{a_se} and automatic engineering of wave packets for molecules or monochrome light fields with a genetic algorithm~\cite{a_ga_prl, a_ga_apb, a_ga_science}. Our approach of quantum automatic control contrasts with the aforementioned methods as we adopt the fundamentals of QIS. In this scheme, information processing tasks are performed by automatically adjusting the parameters of internal unitary operation. Hence we call the entire system including the unitary operation device of the quantum system, the single-shot measurement, and the feedback system as the ``quantum learning machine ({\it QLeM}).''

In this Letter, we investigate the possibility of the QLeM to develop a quantum algorithm. As an example for a deterministic quantum algorithm, we demonstrate that the QLeM learns Deutsch's task and always finds itself a quantum algorithm, which is different from but equivalent to Deutsch's algorithm.

\begin{figure}[b]
\includegraphics[width=0.35\textwidth]{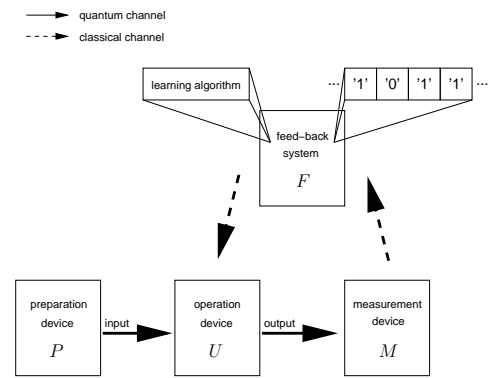}
\caption{Architecture of a quantum learning machine (QLeM), composed of a preparation device $P$, a unitary operation $U$, a single-shot quantum measurement $M$, and a feed-back system $F$ equipped with a classical memory storage $S$.}
\label{fig_qlm}
\end{figure}

{\em Quantum learning machine}.- Given a task by a supervisor, the QLeM learns by itself how to perform the task. A task is represented by a function $f(x)$ where $x$ is an input and $t_x=f(x)$ is the target. Here, we should clarify that $x$ and $t$ are classical numbers and {\it our QLeM takes classical numbers as input values and outputs classical deterministic values}.  However, the internal operations before the measurement are all unitary, which will make sure the advantage of quantum parallelism.  For the function $f$, the supervisor selects a set of $K$ input-target pairs, $T=\{(x_1,f(x_1)), (x_2,f(x_2)), \cdots, (x_K, f(x_K)\}$, and sends the set $T$ to the QLeM through a classical channel. The machine is supposed to learn and to perform the task, now represented by the set $T$.

In order to perform quantum information processing, the QLeM contains a preparation device $P$ to prepare the quantum system $Q$ to be in a certain initial state, an operation device $U$ performing a unitary operation on $Q$, and a measurement device $M$. The QLeM is to provide the basic building blocks, $P$-$U$-$M$, of information processing with a feed-back system $F$, so that $F$ can adjust the control parameters of $U$, depending on the measurement outcome in $M$. For this purpose the feed-back system $F$ is equipped with a classical memory storage $S$, which stores the parameter values of $U$ and records the measurement outcomes at $M$, and two classical channels $C_{FU}$ and $C_{MF}$, where $C_{FU}$ ($C_{MF}$) enables one-way communication from $F$ to $U$ (from $M$ to $F$). Fig.~\ref{fig_qlm} presents the schematic diagram of the QLeM.

The feed-back system $F$ is responsible for QLeM's learning and eventually performing the correct task: It controls the operation device $U$ to eventually bring out the target value for a given input in the set of task $T$. The feed-back system $F$ determines if the target $t$ has been obtained, as monitoring the outcome $m$ from the measurement device $M$. The outcome is transferred from $M$ to $F$ through the classical channel $C_{MF}$.

An input value $x$ can be encoded either in the preparation $P$ or operation $U$ device. In most cases, encoding in $U$ is appropriate and this is the case for finding Deutsch's algorithm as shown later. In order to incorporate the encoding process, $U$ is decomposed into three sub devices $U_1$, $U_2$ and $U_3$, of which the middle one $U_2$ performs an operation according to the input
value $x$.

The QLeM runs in an iterative way as checking if it always works for the given task $T$. At the first iteration, the feed-back system $F$ selects {\it randomly} an element $(x,t_x) \in T$. It prepares the control parameters in the operation device $U$. The action of the sub-device $U_2$ is determined by the input value $x$.  Thus $F$ dials the predetermined value of the parameter $\mathbf{p}_2(x)$ for $U_2$. Then it chooses {\em arbitrary} values of parameters $\mathbf{p}_1$ and $\mathbf{p}_3$ for $U_1$ and $U_3$, respectively. A typical set of such parameters is represented by a unitary operator,
\begin{eqnarray}
\hat{U}(\mathbf{p}) = e^{-i \mathbf{p} \cdot \mathbf{G}},
\end{eqnarray}
where $\mathbf{G}$ is a vector of SU($d$) group generators with $d$ as the dimension of Hilbert space and $\mathbf{p}$ is called a coherent or generalized Bloch vector~\cite{a_ch-vec}. The arbitrariness in choosing parameter values is crucial in our approach of the QLeM, as this implies the machine does not require any {\em a priori} knowledge about the algorithm (Note: predetermining the parameters in $U_2$ is a part of defining the task, not a part of the algorithm). Let us assume that the initial state prepared by $P$ for the quantum system $Q$ is $|0\rangle$ . After going through $U$, $Q$ becomes to be in the output state,
\begin{eqnarray}
|\psi_x\rangle = \hat{U}_x |0\rangle = \hat{U}_3(\mathbf{p}_3) \hat{U}_2(\mathbf{p}_2(x))
\hat{U}_1(\mathbf{p}_1) |0 \rangle.
\label{eq:unitop}
\end{eqnarray}
The device $M$ measures $Q$ in the standard basis $\{|m\rangle\}$ and its outcome $m$ is sent to $F$ through $C_{MF}$. If $m$ is equal to the target $t_x$, $F$ records ``success'', say bit `1', in the classical memory storage $S$ and, otherwise, it records ``failure'', bit `0', in $S$.

\begin{figure}[b]
\includegraphics[width=0.45\textwidth]{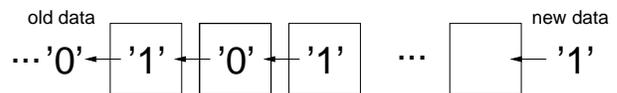}
\caption{The scheme for updating the records in the classical memory
  storage $S$ when $S$ is fully occupied. Here, `1' and `0' denote
  success and failure, respectively. $S$ records sequentially each
  measurement outcome. If it is fully occupied, the oldest data is deleted. Then the remaining data are shifted by one cell into the next
  position and the newly emptied memory cell is filled up with the new data.}\label{fig_st}
\end{figure}

At every iteration, the QLeM repeats the preparation process to re-initialize the quantum state. Then, the feedback system $F$ adjusts the control parameters $\mathbf{p}_{1,3}$ in the operation device $U$. The classical memory storage $S$ keeps the record of success/failure.  If it is fully occupied, $S$ eliminates the oldest record and shifts each record to the next cell, as seen in Fig.~\ref{fig_st}. If the storage is all filled by success, $F$ terminates the learning process, which is called the halt condition, and announces the values of $\mathbf{p}_{1,3}$ to the supervisor. The memory size, denoted by $N$, decides the precision of the QLeM.   We now have a learning
probability $P(n)$ which denotes the probability that the QLeM completes the learning process before or at the $n$th iteration. We can also define a survival probability $Q(n)=1-P(n)$ as the probability that the QLeM does not complete until $n$ (the term survival probability is from the theory of random walks with a trap).

A learning algorithm tells the feedback system $F$ how to update the parameter values $\mathbf{p}_{1,3}$ in the operation device $U$ and when to halt the learning process before announcing $\mathbf{p}_{1,3}$ to the supervisor. Our design of the learning algorithm is as follows.  At the reception of the parameter vector $\mathbf{p}^{(n)}=(\mathbf{p}_1^{(n)}, \mathbf{p}_2^{(n)})$ for the $n$th iteration, (and also $\mathbf{p}_2(x)$ for the random input value $x$), $U$ performs the corresponding unitary operation and results in an output state $|\psi_x\rangle$, as in Eq.~(\ref{eq:unitop}). Measuring $|\psi_x\rangle$, $M$ judges if the measurement outcome is the same as the target $t_x$ and it sends the result to $F$. $F$ records the outcome in $S$ as described earlier. If the operation was successful, $F$ trusts the parameter values and leaves them unchanged: $\mathbf{p}^{(n+1)} = \mathbf{p}^{(n)}$. Otherwise, $F$ needs to modify the parameter values. Instead of using any {\em a priori} knowledge on the algorithm, $F$ generates another {\em random} vector $\mathbf{r}$ and adjusts the parameter vector as
\begin{eqnarray}\label{eq:pm_method}
\mathbf{p}^{(n+1)} = \mathbf{p}^{(n)} + \frac{N_0}{N_T}\mathbf{r},
\end{eqnarray}
where $N_0$ and $N_1$ are respectively the numbers of failure and success events so far, and $N_T = \min(N,N_1+N_0)$. Our learning algorithm is intuitively understandable: the more the number of failure events, the more the adjustment is imposed to the parameter.  $\mathbf{p}$ remains invariant if all events were successful.  Note that the oldest records in $S$ will be eliminated as the learning process continues, keeping $F$ on using the latest records for the adjustment.

It is worth noting that the QLeM completes the learning process much more efficiently than the case of the choice of parameters randomly without an access to the memory of success or failure~\cite{note3}. The QLeM with this learning algorithm can be modeled by a random walk where its survival probability becomes an exponential function in the form of $e^{(n-1)/n_c}$, with a characteristic constant $n_c$. Once the QLeM completes the learning process for the given task, it transmits the parameter values to the supervisor. Then, the supervisor analyzes them, decomposes the unitary operations into a sequence of universal gates~\cite{b_qinfo,a_qa1}, and compares the sequence to a classical one. If it works better than its classical counterpart, the sequence is a quantum algorithm for the task.

\begin{figure}[b]
\includegraphics[width=0.4\textwidth]{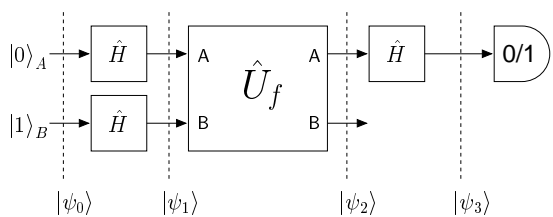}
\caption{Quantum circuit for Deutsch's algorithm, where $\hat{H}$ is Hadamard operator and $\hat{U}_f$ is an operator for a given function $x_i$ which results in the output state $\ket{k_1,k_2\oplus x_i(k_1)}$ when applied on the initial state $\ket{k_1,k_2}$.}\label{fig:da0}
\end{figure}

{\em Example: Finding Deutsch's algorithm}.- To investigate the possibility of QLeM for developing new quantum algorithms, we consider a QLeM for Deutsch's problem of judging if a function is balanced or constant. The QLeM will be shown to find optimal algorithms, possibly different from but equivalent to Deutsch's original algorithm. For the purpose we employ {\em a quantum Monte-Carlo method} to numerically simulate an experiment.

Deutsch's problem is to decide if an arbitrary binary function is constant or balanced~\cite{d_a}. Consider a function $x$ with the domain and the image both being the binary set $\{0,1\}$. There are four possible functions, $x_i$:
\begin{eqnarray}
\label{eq:da0_1}
\begin{matrix}
x_0(0)=0, & x_0(1)=0; & ~~~~ x_1(0)=0, & x_1(1)=1; \\
x_2(0)=1, & x_2(1)=0; & ~~~~ x_3(0)=1, & x_3(1)=1.
\end{matrix}
\end{eqnarray}
If $x_i(0)=x_i(1)$ as in $x_{0,3}$, the function $x_i$ is said to be constant.  Otherwise, the function is balanced (as in $x_{1,2}$). The classical algorithm is simple: Obtaining the values $y=x(k)$ for   $k=0,1$, it judges if $x(0) = x(1)$. Such an algorithm requires two queries of $x$ for the both values of $k$. On the other hand, Deutsch's algorithm enables the judgement of $x$ only by a single query, as it uses a quantum superposition of $k=0$ and $1$.

The quantum circuit for Deutsch's algorithm is presented in Fig.~{\ref{fig:da0}}. In the circuit, ${H}$ is the Hadamard gate which transforms $|0\rangle$ or $|1\rangle$ to a quantum superposition $\hat{H}|0\rangle = (\ket{0}+\ket{1})/\sqrt{2}$ or $\hat{H}|1\rangle =
(\ket{0}-\ket{1})/\sqrt{2}$, respectively. The operation device $U_f$ is a gate to calculate a given function $x_i$, which transforms $|k_1, k_2\rangle_{AB}$ to $|k_1, k_2 \oplus x_i(k_1)\rangle_{AB}$. After going through the gates in the circuit, the qubits $A$ and
$B$ are in the state
\begin{eqnarray}
|\psi_f \rangle = \left\{
\begin{array}{ll}
\pm |0\rangle_A \frac{(|0\rangle - |1\rangle)_B}{\sqrt{2}}, & \textrm{if $x_i$ is constant},\\
\pm |1\rangle_A \frac{(|0\rangle - |1\rangle)_B}{\sqrt{2}}, & \textrm{if $x_i$ is balanced}.
\end{array}
\right.
\end{eqnarray}
The outcome at the measurement device $M$ tells us if $x_i$ is constant or balanced. It needs to detect a single qubit in the standard basis $\{|0\rangle, |1\rangle\}$. The efficiency of the quantum algorithm is dramatically improved by enlarging the domain of a function, as in
Deutsch-Jozsa algorithm~\cite{dj_a1, dj_a2}.

\begin{figure}[b]
\includegraphics[width=0.4\textwidth]{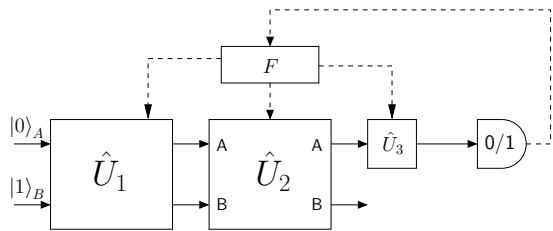}
\caption{Architecture of the QLeM for Deutsch's task, where the unitary operation $U$ consists of three sub-operations (see the text). Here $\hat{U}_1$ and $\hat{U}_3$ are two-qubit and single-qubit operators, respectively.}\label{fig_da2}
\end{figure}

We consider the QLeM that learns Deutsch's task and finds by itself an optimal internal operation.  Deutsch's task is represented by a set,
\begin{eqnarray}
T=&&\{(x_0, f(x_0)=\textrm{c}), (x_1, f(x_1)=\textrm{b}),
\nonumber \\
&&~~(x_2, f(x_2)=\textrm{b}), (x_3, f(x_3)=\textrm{c})\},
\end{eqnarray}
where the input $x_i$ is defined in Eq.~(\ref{eq:da0_1}) and `c' and `b' stand for constant and balanced, respectively.

The QLeM is schematically presented in Fig.~\ref{fig_da2}. The preparation device prepares two qubits $A$ and $B$ to be in certain fixed states, say $|0\rangle$ and $|1\rangle$, respectively.  In order to maximize the quantum parallelism, the number of input qubits has been chosen to be 2, for $k$ can take two values 0 and 1 in $x_i(k)$.  The choice of $|0\rangle$ and $|1\rangle$ for $A$ and $B$ can be random.  They may be chosen differently as far as they are fixed throughout the learning process.  The middle sub-device $U_2$ is placed to calculate the function of a given input $x_i(k)$.  The first sub-device $U_1$ performs a two-qubit unitary operations. On the other hand, $U_3$ does a single-qubit unitary operation before the measurement.  The single qubit measurement has been chosen as there is only one bit of information, c and b, for the target value.  $U$ has 18 control parameters \footnote{There are $d^2-1$ number of paramters for SU($d$) operations.}; $4^2-1=15$ for two-qubit operation $U_1$ and $2^2-1=3$ for one-qubit operation $U_3$. The sub-device $U_2$ is a part of defining the task and the parameter values are predetermined with respect to the input function $x_i$, such that $U_2$ transforms an input state $|k_1, k_2\rangle_{AB}$ to $|k_1, k_2 \oplus
x_i(k_1)\rangle_{AB}$.

\begin{figure}[t]
\includegraphics[angle=270,width=0.23\textwidth]{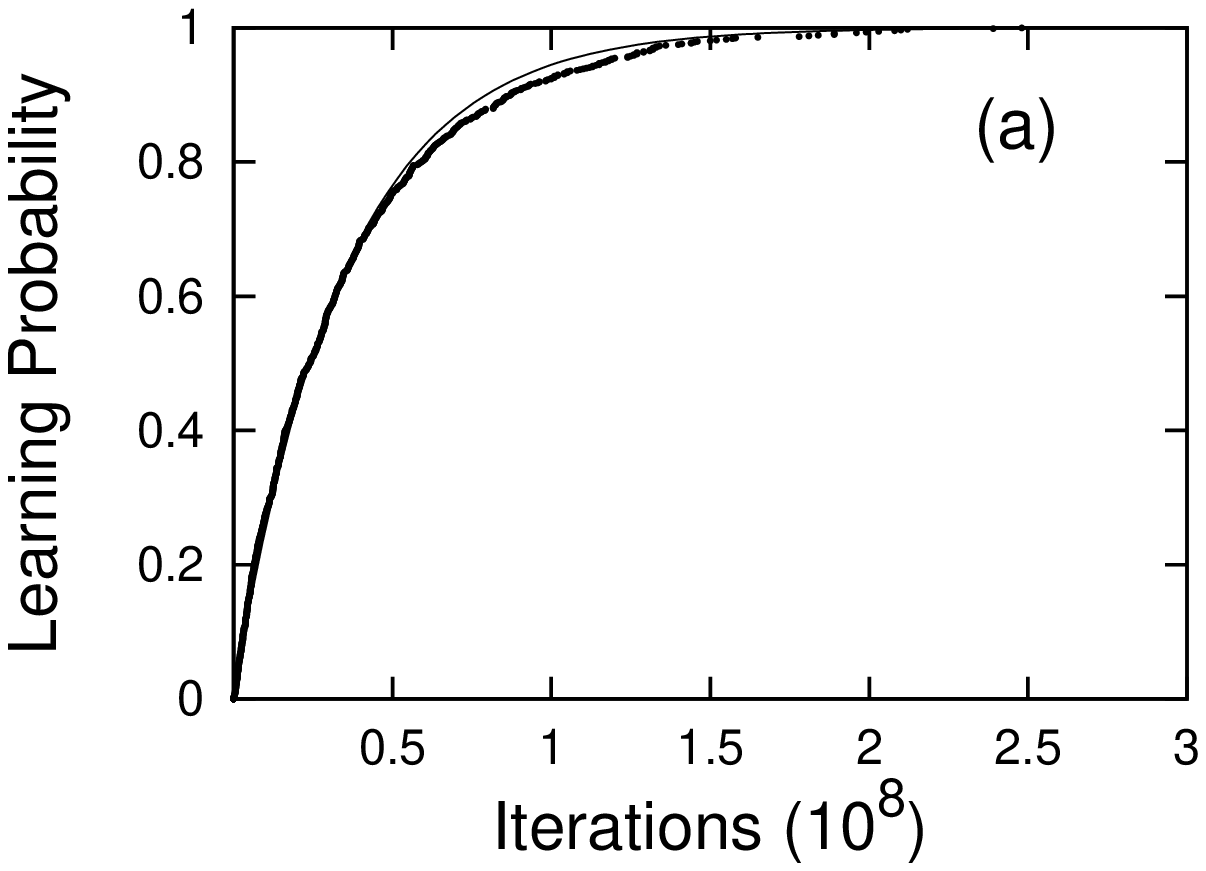}
\includegraphics[angle=270,width=0.23\textwidth]{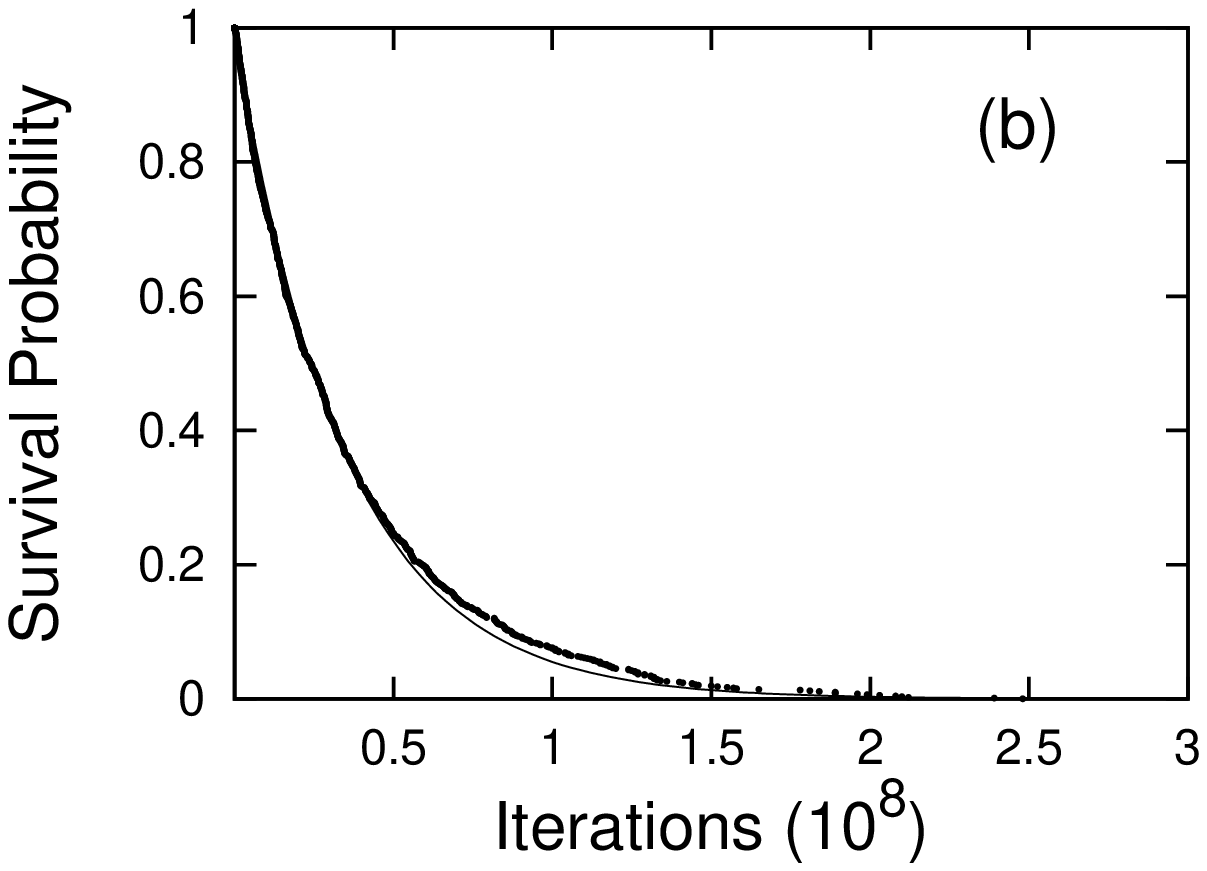}
\caption{(a) Learning probability $P(n)$ and (b) survival probability
  $Q(n)=1-P(n)$ in terms of the number of iterations $n$, constructed by
  quantum Monte-Carlo method with 1000 trials for 300 bits of the
  classical memory storage. Solid line is a fitting function.  The
  survival probability is well fitted to $Q(n) = e^{-(n-1)/n_c}$ with a
  fitting parameter $n_c$.}
\label{grp:da2}
\end{figure}

We performs the quantum Monte-Carlo simulation, sampling 1000 trials, where the QLeM resets all the parameters for each trial. In Fig.~\ref{grp:da2}(a), we present a learning probability $P(n)$, constructed by 1000 trials with 300 bits of the classical memory employed. The survival probability $Q(n)=1-P(n)$ in Fig.~\ref{grp:da2}(b) is well fitted to an exponential function, $e^{-(n-1)/n_c}$ with the characteristic constant $n_c$, which characterizes how many iterations are necessary for the completion of learning. Note that $n_c$ is finite and smaller than that for the case of learning algorithm without a memory facility~\cite{note3}. The QLeM is able to learn how to perform Deutsch's task in a finite number of iterations, which is witnessed by the exponential decay of the survival probability with regard to the number of iterations. We have also noted that each unitary operator $\hat{U}_1$ after completing the learning process is a product of two one-qubit operators within the precision allowed. The two one-qubit operators are in general not equal to the two Hadamard operators as in Deutsch's algorithm. Nevertheless,the internal operations, $\hat{U}_1$ and $\hat{U}_3$ perform Deutsch's task so that the quantum algorithms identified by the QLeM are equivalent to the Deutsch's original one as far as all the one-qubit operations cost the same.

\begin{figure}[t]
\includegraphics[angle=270,width=0.23\textwidth]{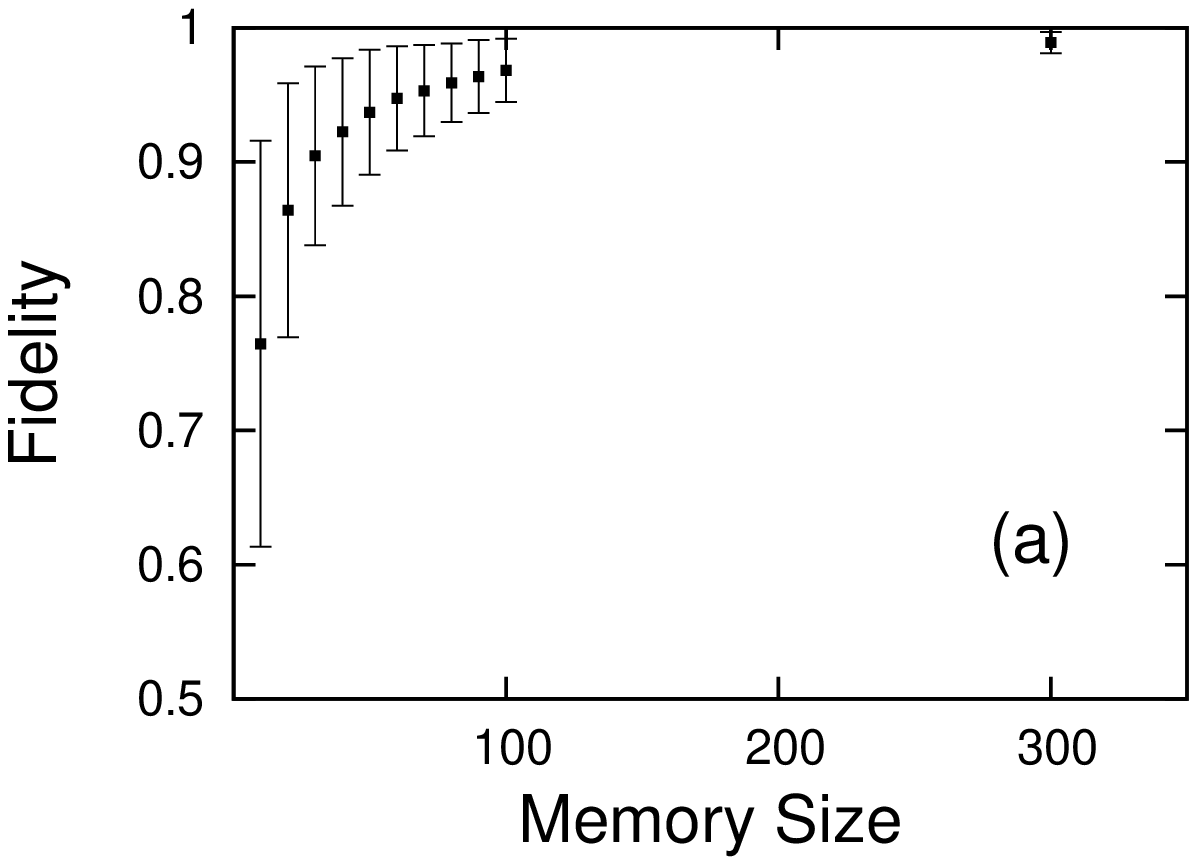}
\includegraphics[angle=270,width=0.23\textwidth]{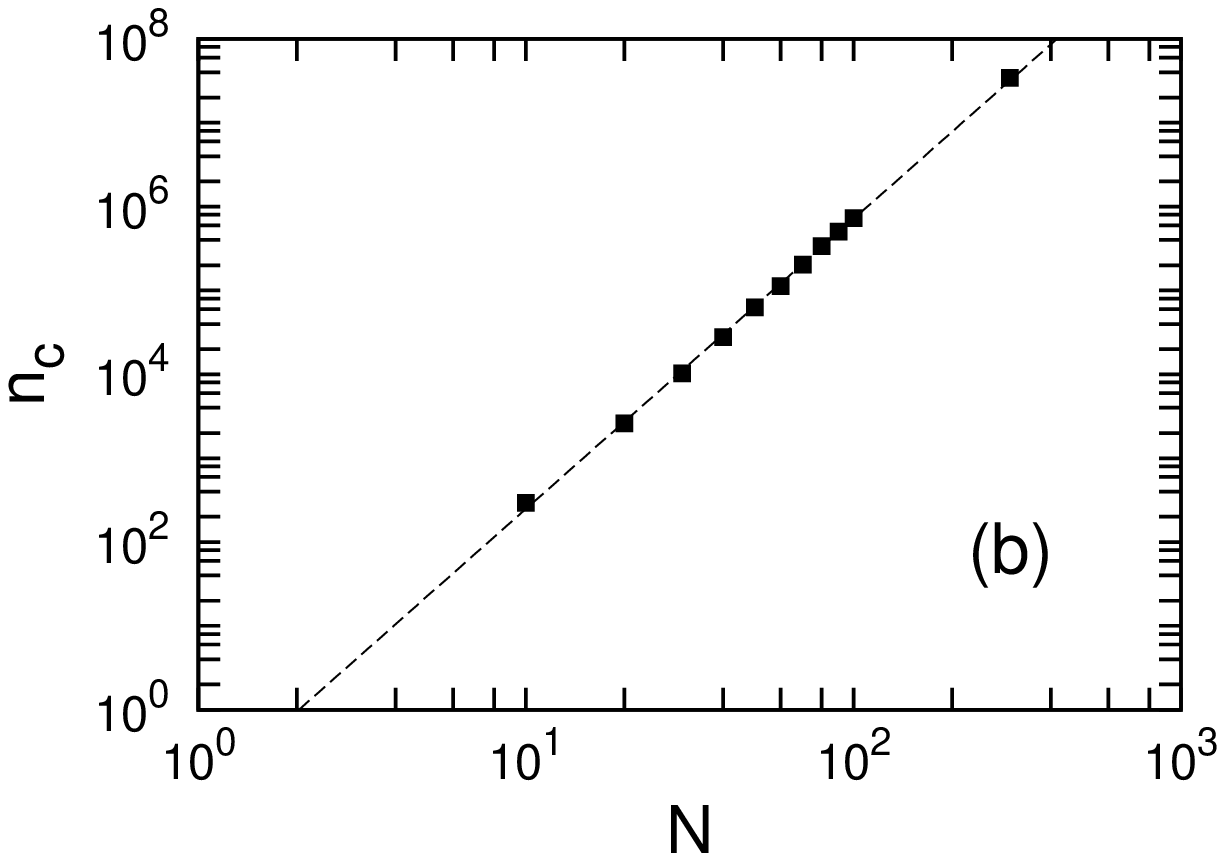}
\caption{a) Fidelity (${\cal F}$) averaged over 1000 trials for a given memory size $N$. Each error bar is a standard deviation over the trials. (b) Characteristic constant $n_c$ as a function of $N$ on a log-log scale.  The data are well fitted to $n_c = A N^D$ with $D \simeq 3.46$ and $A \simeq 10^{-1.06}$.}
\label{fig:fidelity}
\end{figure}

We calculate the fidelity to find out how close the quantum state before the measurement is to the target state $|t_x\rangle$ when the QLeM completes the learning. In Fig.~\ref{fig:fidelity}(a), we present the average fidelity with respect to a memory size $N$. Here the fidelity ${\cal F}=\sum_{x=0}^3 \langle
t_x| \hat{\rho}_x |t_x \rangle/4$ is averaged over 1000 trials for a given $N$, where  $\hat{\rho}_x$ is the density operator for the QLeM output state. The average fidelity ${\cal F}$ approaches unity as the memory size $N$ increases, as we expected. For instance, ${\cal F}$ is as large as 0.989 for $N=300$.  Fig.~\ref{fig:fidelity}(b) presents the characteristic constant $n_c$ as a function of $N$ on a log-log scale. It is found that the characteristic constant $n_c$ is well fitted to a polynomial function of $N$, i.e., $n_c = A N^D$ with $D\simeq 3.46$ and $A \simeq 10^{-1.06}$.

In summary, we have proposed a novel notion of the QLeM for automatically controlling quantum coherence to find a quantum algorithm by itself. We have illustrated, for Deutsch's task, how the QLeM finds a quantum algorithm and shown that the algorithms identified by the QLeM are equivalent to Deutsch's algorithm using the fidelity which is $\simeq 1$ for a finite number of iterations. This will open a new field of research to find a new quantum algorithm and further studies are necessary to improve the learning algorithm.

\acknowledgements
We thank Prof. \v{C}. Brukner for discussions. We acknowledge financial support from Korean Research Foundation Grant funded by the Korean Government (KRF-2005-041-c00197), UK EPSRC and QIP IRC.

{\em Note added.}- As completing this work, we recently found a related work, which considers probabilistic quantum learning for database search and factorization tasks~\cite{note2}.

\bibliography{common}

\begin{thebibliography}{17}
\expandafter\ifx\csname natexlab\endcsname\relax\def\natexlab#1{#1}\fi
\expandafter\ifx\csname bibnamefont\endcsname\relax
  \def\bibnamefont#1{#1}\fi
\expandafter\ifx\csname bibfnamefont\endcsname\relax
  \def\bibfnamefont#1{#1}\fi
\expandafter\ifx\csname citenamefont\endcsname\relax
  \def\citenamefont#1{#1}\fi
\expandafter\ifx\csname url\endcsname\relax
  \def\url#1{\texttt{#1}}\fi
\expandafter\ifx\csname urlprefix\endcsname\relax\def\urlprefix{URL }\fi
\providecommand{\bibinfo}[2]{#2}
\providecommand{\eprint}[2][]{\url{#2}}

\bibitem[{\citenamefont{Feynman}(1986)}]{feynman}
\bibinfo{author}{\bibfnamefont{R.}~\bibnamefont{Feynman}},
  \bibinfo{journal}{Opt. News.} \textbf{\bibinfo{volume}{11}},
  \bibinfo{pages}{11} (\bibinfo{year}{1986}).

\bibitem[{\citenamefont{Deutsch and Jozsa}(1992)}]{dj_a1}
\bibinfo{author}{\bibfnamefont{D.}~\bibnamefont{Deutsch}} \bibnamefont{and}
  \bibinfo{author}{\bibfnamefont{R.}~\bibnamefont{Jozsa}},
  \bibinfo{journal}{Proc. R. Soc. London A} \textbf{\bibinfo{volume}{439}},
  \bibinfo{pages}{553} (\bibinfo{year}{1992}).

\bibitem[{\citenamefont{Cleve et~al.}(1998)\citenamefont{Cleve, Ekert,
  Macchiavello, and Mosca}}]{dj_a2}
\bibinfo{author}{\bibfnamefont{R.}~\bibnamefont{Cleve}},
  \bibinfo{author}{\bibfnamefont{A.}~\bibnamefont{Ekert}},
  \bibinfo{author}{\bibfnamefont{C.}~\bibnamefont{Macchiavello}},
  \bibnamefont{and} \bibinfo{author}{\bibfnamefont{M.}~\bibnamefont{Mosca}},
  \bibinfo{journal}{Proc. R. Soc. London A} \textbf{\bibinfo{volume}{454}},
  \bibinfo{pages}{339} (\bibinfo{year}{1998}).

\bibitem[{\citenamefont{Shor}(1997)}]{a_qa2-Shor}
\bibinfo{author}{\bibfnamefont{P.~W.} \bibnamefont{Shor}},
  \bibinfo{journal}{SIAM J. comput.} \textbf{\bibinfo{volume}{26}},
  \bibinfo{pages}{1484} (\bibinfo{year}{1997}).

\bibitem[{\citenamefont{Ekert and Jozsa}(1996)}]{a_qa4-SG}
\bibinfo{author}{\bibfnamefont{A.}~\bibnamefont{Ekert}} \bibnamefont{and}
  \bibinfo{author}{\bibfnamefont{R.}~\bibnamefont{Jozsa}},
  \bibinfo{journal}{Rev. Mod. Phys.} \textbf{\bibinfo{volume}{68}},
  \bibinfo{pages}{733} (\bibinfo{year}{1996}).

\bibitem[{\citenamefont{Grover}(1997)}]{a_qa3-Grover}
\bibinfo{author}{\bibfnamefont{L.~K.} \bibnamefont{Grover}},
  \bibinfo{journal}{Phys. Rev. Lett.} \textbf{\bibinfo{volume}{79}},
  \bibinfo{pages}{325} (\bibinfo{year}{1997}).

\bibitem[{\citenamefont{Nielsen and Chuang}(1999)}]{b_qinfo}
\bibinfo{author}{\bibfnamefont{M.~A.} \bibnamefont{Nielsen}} \bibnamefont{and}
  \bibinfo{author}{\bibfnamefont{I.~L.} \bibnamefont{Chuang}},
  \emph{\bibinfo{title}{Quantum Computation and Quantum Information}}
  (\bibinfo{publisher}{Springer}, \bibinfo{year}{1999}).

\bibitem[{\citenamefont{Purushothaman and Karayiannis}(1997)}]{a_nn}
\bibinfo{author}{\bibfnamefont{G.}~\bibnamefont{Purushothaman}}
  \bibnamefont{and} \bibinfo{author}{\bibfnamefont{N.~B.}
  \bibnamefont{Karayiannis}}, \bibinfo{journal}{IEEE Trans. Neural Networks}
  \textbf{\bibinfo{volume}{8}}, \bibinfo{pages}{679} (\bibinfo{year}{1997}).

\bibitem[{\citenamefont{Fischer et~al.}(2000)\citenamefont{Fischer, Kienle, and
  Freyberger}}]{a_se}
\bibinfo{author}{\bibfnamefont{D.~G.} \bibnamefont{Fischer}},
  \bibinfo{author}{\bibfnamefont{S.~H.} \bibnamefont{Kienle}},
  \bibnamefont{and}
  \bibinfo{author}{\bibfnamefont{M.}~\bibnamefont{Freyberger}},
  \bibinfo{journal}{Phys. Rev. A} \textbf{\bibinfo{volume}{61}},
  \bibinfo{pages}{032306} (\bibinfo{year}{2000}).

\bibitem[{\citenamefont{Judson and Rabitz}(1992)}]{a_ga_prl}
\bibinfo{author}{\bibfnamefont{R.~S.} \bibnamefont{Judson}} \bibnamefont{and}
  \bibinfo{author}{\bibfnamefont{H.}~\bibnamefont{Rabitz}},
  \bibinfo{journal}{Phys. Rev. Lett.} \textbf{\bibinfo{volume}{68}},
  \bibinfo{pages}{1500} (\bibinfo{year}{1992}).

\bibitem[{\citenamefont{Baumert et~al.}(1997)\citenamefont{Baumert, Brixner,
  Seyfried, Strehle, and Gerber}}]{a_ga_apb}
\bibinfo{author}{\bibfnamefont{T.}~\bibnamefont{Baumert}},
  \bibinfo{author}{\bibfnamefont{T.}~\bibnamefont{Brixner}},
  \bibinfo{author}{\bibfnamefont{V.}~\bibnamefont{Seyfried}},
  \bibinfo{author}{\bibfnamefont{M.}~\bibnamefont{Strehle}}, \bibnamefont{and}
  \bibinfo{author}{\bibfnamefont{G.}~\bibnamefont{Gerber}},
  \bibinfo{journal}{Appl. Phys. B} \textbf{\bibinfo{volume}{65}},
  \bibinfo{pages}{779} (\bibinfo{year}{1997}).

\bibitem[{\citenamefont{Assion et~al.}(1998)\citenamefont{Assion, Baumert,
  Bergt, Brixner, Kiefer, Seyfried, Strehle, and Gerber}}]{a_ga_science}
\bibinfo{author}{\bibfnamefont{A.}~\bibnamefont{Assion}},
  \bibinfo{author}{\bibfnamefont{T.}~\bibnamefont{Baumert}},
  \bibinfo{author}{\bibfnamefont{M.}~\bibnamefont{Bergt}},
  \bibinfo{author}{\bibfnamefont{T.}~\bibnamefont{Brixner}},
  \bibinfo{author}{\bibfnamefont{B.}~\bibnamefont{Kiefer}},
  \bibinfo{author}{\bibfnamefont{V.}~\bibnamefont{Seyfried}},
  \bibinfo{author}{\bibfnamefont{M.}~\bibnamefont{Strehle}}, \bibnamefont{and}
  \bibinfo{author}{\bibfnamefont{G.}~\bibnamefont{Gerber}},
  \bibinfo{journal}{Science} \textbf{\bibinfo{volume}{282}},
  \bibinfo{pages}{919} (\bibinfo{year}{1998}).

\bibitem[{\citenamefont{Hioe and Eberly}(1981)}]{a_ch-vec}
\bibinfo{author}{\bibfnamefont{F.~T.} \bibnamefont{Hioe}} \bibnamefont{and}
  \bibinfo{author}{\bibfnamefont{J.~H.} \bibnamefont{Eberly}},
  \bibinfo{journal}{Phys. Rev. Lett.} \textbf{\bibinfo{volume}{47}},
  \bibinfo{pages}{838} (\bibinfo{year}{1981}).

\bibitem[{\citenamefont{Ryu et~al.}(2008)\citenamefont{Ryu, Bang, and
  Lee}}]{note3}
\bibinfo{author}{\bibfnamefont{S.}~\bibnamefont{Ryu}},
  \bibinfo{author}{\bibfnamefont{J.}~\bibnamefont{Bang}}, \bibnamefont{and}
  \bibinfo{author}{\bibfnamefont{J.}~\bibnamefont{Lee}}, \bibinfo{journal}{in
  preparation}  (\bibinfo{year}{2008}).

\bibitem[{\citenamefont{Barenco et~al.}(1995)\citenamefont{Barenco, Bennett,
  Cleve, DiVincenzo, Margolus, Shor, Sleator, Smolin, and Weinfurter}}]{a_qa1}
\bibinfo{author}{\bibfnamefont{A.}~\bibnamefont{Barenco}},
  \bibinfo{author}{\bibfnamefont{C.~H.} \bibnamefont{Bennett}},
  \bibinfo{author}{\bibfnamefont{R.}~\bibnamefont{Cleve}},
  \bibinfo{author}{\bibfnamefont{D.~P.} \bibnamefont{DiVincenzo}},
  \bibinfo{author}{\bibfnamefont{N.}~\bibnamefont{Margolus}},
  \bibinfo{author}{\bibfnamefont{P.}~\bibnamefont{Shor}},
  \bibinfo{author}{\bibfnamefont{T.}~\bibnamefont{Sleator}},
  \bibinfo{author}{\bibfnamefont{J.~A.} \bibnamefont{Smolin}},
  \bibnamefont{and}
  \bibinfo{author}{\bibfnamefont{H.}~\bibnamefont{Weinfurter}},
  \bibinfo{journal}{Phys. Rev. A} \textbf{\bibinfo{volume}{52}},
  \bibinfo{pages}{3457} (\bibinfo{year}{1995}).

\bibitem[{\citenamefont{Deutsch}(1985)}]{d_a}
\bibinfo{author}{\bibfnamefont{D.}~\bibnamefont{Deutsch}},
  \bibinfo{journal}{Proc. R. Soc. London A} \textbf{\bibinfo{volume}{400}},
  \bibinfo{pages}{97} (\bibinfo{year}{1985}).

\bibitem[{\citenamefont{Gammelmark and Molmer}(2008)}]{note2}
\bibinfo{author}{\bibfnamefont{S.}~\bibnamefont{Gammelmark}} \bibnamefont{and}
  \bibinfo{author}{\bibfnamefont{K.}~\bibnamefont{Molmer}},
  \bibinfo{journal}{arXiv:0803.1418}  (\bibinfo{year}{2008}).

\end{thebibliography}

\end{document}